\documentstyle[twocolumn,aps,epsf]{revtex}
\begin{document}
\draft

\newcommand{\mytitle}[1]{
\twocolumn[\hsize\textwidth\columnwidth\hsize
\csname@twocolumnfalse\endcsname #1 \vspace{1mm}]}

\mytitle{
\title{Separately contacted edge states:\\
A spectroscopic tool for the investigation of the quantum Hall
effect}
\author{A. W\"urtz$^{a,b}$, R. Wildfeuer$^b$, A. Lorke$^{a,b}$}
\address{$^a$Laboratorium f\"ur Festk\"orperphysik, Gerhard-Mercator-Universit\"at, \\
Lotharstr. 17, D-47048 Duisburg, Germany \\
$^b$Sektion Physik and Center for NanoScience,
Ludwig-Maximilians-Universit\"at M\"unchen, Geschwister-Scholl-Platz 1,
D-80539 M\"unchen, Germany}
\author{ E. V. Deviatov, V. T. Dolgopolov}
\address{Institute of Solid State Physics, Chernogolovka, Moscow
District 142432, Russia}
\date{\today}
\maketitle

\begin{abstract}

 Using an innovative combination of a quasi-Corbino
sample geometry and the cross-gate technique, we have developed a
method that enables us to separately contact single edge channels
in the quantum Hall regime and investigate equilibration among
them. Performing 4-point resistance measurements, we directly
obtain information on the energetic and geometric structure of the
edge region and the equilibration-length for current transport
across the Landau- as well as the spin-gap. Based on an almost
free choice in the number of participating edge channels and their
interaction-length a systematic investigation of the
parameter-space becomes possible.
\end{abstract}
\pacs{}}

\section{Introduction} \label{sec:introduction}

The concept of one-dimensional current-carrying edge channels in
the quantum Hall regime has been the subject of detailed
experimental and theoretical studies in recent years. It is now
widely accepted as a basic ingredient for the understanding of the
quantum Hall effect
\cite{Halperin82,Haug93,MacDonald84,Buttiker85,Buttiker88,Chklovskii92,Chklovskii93}.
In the edge state picture, the resistance quantization is
attributed to current transport in quasi one-dimensional edge
channels along the sample boundaries, assuming negligible
backscattering between opposing edges of the two-dimensional
electron gas (2DEG) due to vanishing bulk conductivity
\cite{Buttiker85,Buttiker88}. Edge channels are formed at the
intersection of Landau-levels with the Fermi-energy, where the
presence of unoccupied states allows for current transport.
Theoretical treatments, such as the compressible/incompressible
liquid picture \cite{Chklovskii92,Chklovskii93} have led to a
better understanding of the energetic structure of the 2DEG-edge.
Various efforts have been made to experimentally investigate the
details of edge-reconstruction and edge channel transport by means
of tunneling- \cite{Zhitenev95,Hwang93} or
capacitance-spectroscopy \cite{Takaoka94}. At the same time
advances in far infrared- \cite{Merz93,Lorke96,Hirakawa96},
time-resolved transport- \cite{Zhitenev93} and
edge-magnetoplasmon-spectroscopy \cite{Talyanskii92}, as well as
SET-measurements \cite{Wei98} have given further insight in the
edge structure of the 2DEG. However, the experimental
investigation of the potential profile at the sample boundaries
remains a challenging task.

Here we report on the development and application of a new sample
geometry, which renders possible a direct measurement of the
charge transfer between adjacent edge channels at the sample
boundaries. The combination of a quasi-Corbino \cite{Oto99}
topology and the cross-gate technique \cite{Haug88,Washburn88} has
a number of advantages, which make it a versatile tool for
studying edge channel transport: Requiring no sophisticated sample
fabrication such as high-resolution lithography or cleaved-edge
overgrowth \cite{Hilke01}, it offers true 4-probe measurements in
both, the linear and the non-linear transport regime. Furthermore
it can ``dissect'' the edge channel structure, i.e. for $n$ edge
channels, anyone of the $n-1$ gaps can individually be addressed.
This is done by separately contacting single edge channels,
selectively populating them \cite{vanWees89,Komiyama89} and
bringing them into a controlled interaction. We present
experimental data obtained using this geometry which allows us to
determine the equilibration-length in the linear regime and get
information on the energetic edge structure (i.e. the spin- and
Landau-gaps) by non-linear I--V-spectroscopy.

This report is organized as follows: Section \ref{sec:geometry}
gives a detailed description of the sample geometry, and the
general concept behind the experiment. Also the application of the
Landauer-B\"uttiker formalism \cite{Buttiker85,Buttiker88} to the
given topology is briefly outlined. Experimental results for
different filling factor combinations, temperatures and
interaction-lengths are presented in Section
\ref{sec:experimental}. These will be discussed in terms of the
edge-reconstruction picture and evaluated with regard to the spin-
and Landau-gap structure.

\section{Device geometry}\label{sec:geometry}
\subsection{Concept and realization}\label{sec:concept}

The samples were fabricated from a molecular beam
epitactically-grown GaAs/AlGaAs heterostructure, consisting of a 1
$\mu$m GaAs buffer layer, 20 nm undoped Al$_{0.3}$Ga$_{0.7}$As, 25
nm silicon-doped Al$_{0.3}$Ga$_{0.7}$As covered by a 45 nm
superlattice and cap-layer. The mobility at liquid Helium
temperature was 800 000 cm$^{2}$/Vs and the carrier density 3.7
$\cdot 10^{11}
 $cm$^{-2}$. All structures were patterned by standard
photolithography. The mesa was defined by wet chemical etching.
The gate-electrode consists of 5 nm thermally evaporated NiCr
(1:1). Ohmic contacts with typical resistances of 200 $\Omega$ at
30 mK are provided by alloyed AuGe/Ni/AuGe (88:12) pads.

\vspace*{-.3cm}
\begin{center}
\begin{figure}
\epsfsize=0.75 \columnwidth \epsfbox{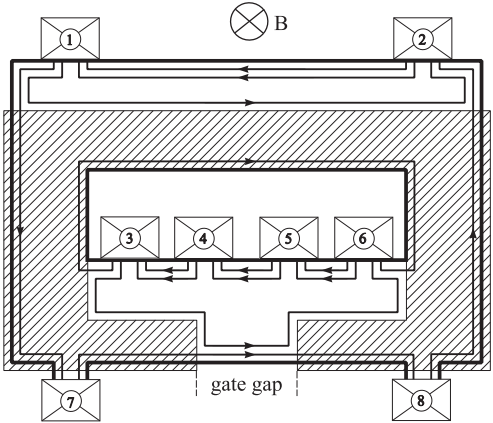} \vspace*{.2cm}
\caption{Schematic diagram of the device geometry. Contacts are
positioned along the inner and outer edge of the ring-shaped mesa.
The shaded area represents the Schottky-gate. Arrows indicate the
direction of electron drift in the edge channels for the outlined
configuration: filling factor $\nu=2$ in the ungated regions and
$g=1$ under the gate.} \label{fig1}
\end{figure}
\end{center}

For the measurements reported here a total of 4 samples with
slightly different geometries, fabricated from the same
heterostructure, were investigated. The device geometry is given
in Fig.~\ref{fig1}, where the ring-shaped mesa is shown by the
thick outline and the gate-electrode is indicated by the hatched
area. In a quantizing magnetic field and for zero gate voltage,
the present quasi-Corbino geometry has two sets of edge channels,
called ``inner" and ``outer" edge states in the following. The
shape and location of the gate is chosen such that, by application
of a suitable negative bias, the inner edge channels can be
redirected to run along the outer edge in the gate-gap region.
This situation is shown in Fig.~\ref{fig1} for equilibrium
conditions, i.e. without an external applied current. Here, the
filling factors have been adjusted to $\nu=2$ in the ungated
regions and $g=1$ under the gate. In the gate-gap there are two
adjacent channels running in parallel. One of them is reflected by
the gate potential, and connected only to the inner Ohmic
contacts. The other, outer edge channel continues to run along the
etched boundary, even under the gate-electrode, and is therefore
connected only to the contacts positioned along the outer mesa
edge. When the bulk region of the 2DEG under the gate is in its
insulating state (integer filling factor and sufficiently low
temperatures) current transport between inner and outer Ohmic
contacts is possible only by charge equilibration among
neighboring channels in the gate-gap region. The present geometry
therefore allows for a direct investigation --- in 2- or 4-probe
geometries --- of transport across the incompressible strip that
separates the compressible edge channels within the gate-gap.

\vspace*{-.3cm}
\begin{center}
\begin{figure}
\epsfsize=1 \columnwidth \epsfbox{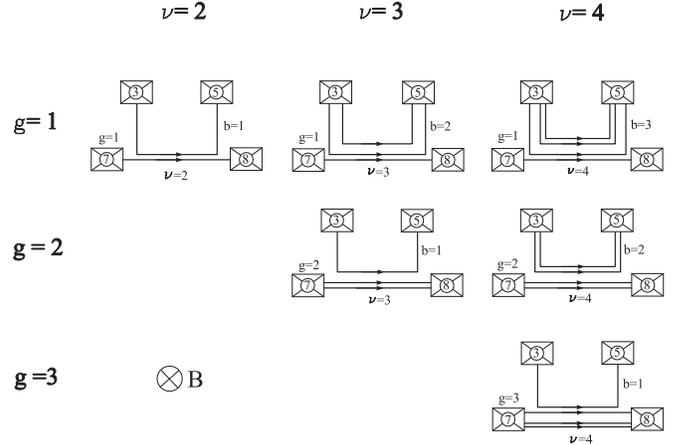} \vspace*{.1cm}
\caption{Survey on the adjusted filling factor combinations in the
IQHE regime. Outlined are only the traces of edge channels in the
gate-gap region. Here $\nu$ indicates the number of occupied
spinpolarized edge states in the ungated regions, $g$ the number
of channels transmitted under the gate and $b$ the number of
channels reflected by the gate potential.} \label{fig2}
\end{figure}
\end{center}

Figure~\ref{fig2} illustrates the versatility of the present
geometry. For any given integer filling factor $\nu$ all
incompressible strips can be probed by adjusting the gate voltage
to the appropriate integer filling factor $g$ in the gated region.
When spin-splitting is resolved and the filling factors are
adjusted to $\nu=4$ and $g=1$, the outermost spin-induced
incompressible strip can be probed, whereas for $\nu=4$ and $g=2$
the wider, Landau-gap-induced strip between the second and third
edge channel can be investigated. For clarity, in Fig.~\ref{fig2}
only the four contacts used as current leads or voltage probes in
the experiment are shown. Furthermore, only edge channels running
by the interaction region (i.e. the gate gap) are drawn. In the
evaluation of the experimental data it is of course necessary to
take into account the complete sample geometry, as it will be
discussed in the following section.

\subsection{Application of the Landauer-B\"uttiker
formalism to the experimental setup} \label{sec:LBapplic}

Despite the apparent complexity of the device geometry, a relation
between the measured resistances and equilibration based on
inter-edge-channel transport (expressed in terms of an effective
transmission-coefficient $T$ across the incompressible strip) can
be derived in a straightforward manner. In the following, we will
treat the case of a 4-probe measurement, restricting the
discussion to the leads 3, 5, 7, 8. It is easy to show that the
presence of the unused (floating) contacts 1, 2, 4, 6 neither
changes any of the considerations below nor the obtained
experimental results. Based upon the absence of backscattering
across the bulk and the conservation of current, a
Landauer-B\"uttiker-type multichannel-multiprobe formula can be
derived (see Appendix A) to give the following results for the
four different contact configurations for measuring the 4-probe
resistance between inner and outer edge channels:

\begin{eqnarray}
R_{73,85}&=&\frac{h}{e^{2}}\left[\frac{1}{T}-\frac{\nu+g }{\nu
g}\right]\nonumber\\
R_{75,83}&=&\frac{h}{e^{2}}\left[\frac{1}{T}-\frac{1}{g}\right]\nonumber\\
R_{85,73}&=&\frac{h}{e^{2}}\left[\frac{1}{T}\right]\nonumber\\
R_{83,75}&=&\frac{h}{e^{2}}\left[\frac{1}{T}-\frac{1}{\nu}\right]
\label{eq:Req}
\end{eqnarray}

Here $R_{ij,kl}$ is the resistance measured between the current
leads $i$ and $j$ and voltage probes $k$ and $l$. Note that
exchanging current and voltage probes changes the value of the
resistance because of the chirality of edge states ($R_{ij,kl}(B)
= R_{kl,ij}(-B) \neq R_{kl,ij}(B)$). The measurements can
therefore be used to confirm the direction of electron-drift,
respectively the orientation of the magnetic field in the
experimental setup.

Combining equation \ref{eq:Req} with the relation between $T$ and
the equilibration-length \cite{Muller92} $l_{eq}$ gives for the
interaction of two edge channels

\begin{equation}
R_{85,73}=\frac{h}{e^{2}}\left[1+exp\left(\frac{-2d}{l_{eq}}\right)\right]^{-1}
\label{eq:leq}
\end{equation}

with $d$ being the interaction-length, i.e. the gate-gap width.
This means that the present topology makes it possible to
determine the equilibration-length between neighboring edge
channels by a single resistance measurement.

For elevated temperatures or macroscopic interaction-lengths, full
equilibration is expected. Then the transmission-coefficient can
be derived from a summation over all participating edge states and
contacts:

\begin{equation}
T=\sum_{i=1}^{g}\sum_{j=g+1}^{\nu}1/ \nu=(\nu-g)\cdot g/ \nu .
\label{eq:Tfull}
\end{equation}

In this case all the resistances $R_{ij,kl}$ are readily calculated
and can be compared with the experimental values.

Finally, it should be mentioned that at no point in the derivation
of equation \ref{eq:Req} the transmission-coefficient is assumed
to be independent of the electrochemical potential difference
across the incompressible strip. Equilibration between edge states
can therefore be studied as a function of bias. As will be shown
in Section ~\ref{sec:weak}, this I--V-spectroscopy can be used not
only in the linear transport regime ($T$ = const.), but also to
investigate the local energetic structure of the edge region.

\section{Experimental}\label{sec:experimental}

\subsection{Equilibration across the Landau-gap at
liquid helium-temperature}\label{sec:fulleq}

\begin{center}
\begin{figure}
\epsfsize=0.9 \columnwidth \epsfbox{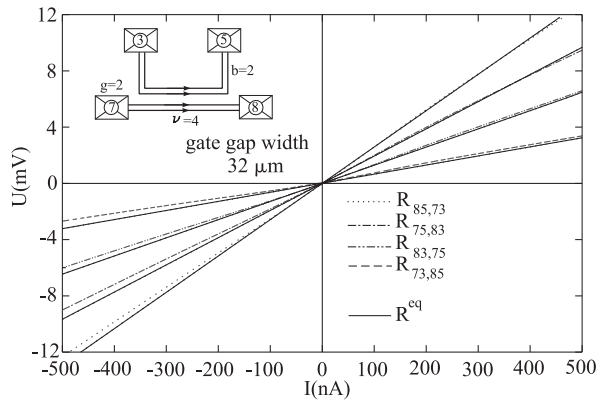} \vspace*{.3cm}
\caption{Measurements at 4.2 K and B = 3.85 T on a device with
gate-gap width 32 $\mu$m for the adjusted filling factor
combination $\nu=4$ and $g=2$. The solid lines indicate the
calculated resistances, according to the equations \ref{eq:Req}
and \ref{eq:Tfull} for fully equilibrated transport.} \label{fig3}
\end{figure}
\end{center}

Figure~\ref{fig3} shows current-voltage curves obtained for a
sample with 32 $\mu$m gate-gap width, a filling factor combination
$\nu=4$, $g=2$ and a temperature of 4.2 K. The inset displays the
corresponding, simplified configuration of edge channels (see also
the discussion of Fig.~\ref{fig2}). As expected from equation
\ref{eq:Req}, each configuration gives a different 4-point
resistance (dashed lines indicate experimental curves). For
comparison, the solid lines show the resistances for complete
equilibration ($T = 1$), calculated as discussed in
Section~\ref{sec:LBapplic}\cite{byline}.

\vspace*{-.3cm}

 The very good agreement
between the calculated resistances for fully equilibrated
transport and the measurements at 4.2 K indicates that at this
temperature the equilibration-length is much shorter than the
gate-gap width of $d=32 \mu$m. Furthermore, the fact that
equations \ref{eq:Req} and \ref{eq:Tfull} can so well account for
the experimental data shows that the discussed Landauer-B\"uttiker
picture is valid and in particular that transport through the bulk
of the 2DEG is negligible. The absence of bulk leakage at 4.2 K
for even filling factors is also confirmed by a direct
determination of the bulk conductivity, adjusting the filling
factors to $\nu=2$, $g=2$, and investigation of a reference-sample
without a gate-gap~\cite{Wildfeuer99}. Also the influence of
non-ideal contact properties proves to be irrelevant.

\subsection{Weak coupling among edge states at 30
mK}\label{sec:weak}

When the samples are cooled down to 30 mK in the mixing chamber of
a $^{3}$He$^{4}$He dilution-refrigerator, the resistance at low
bias increases dramatically. Even for an interaction-length of
more than 30 $\mu$m, resistances of around 10 M$\Omega$ are
observed, corresponding to macroscopic equilibration-lengths (see
section \ref{sec:evaluation}).

\begin{center}
\begin{figure}
\epsfsize=0.97 \columnwidth \epsfbox{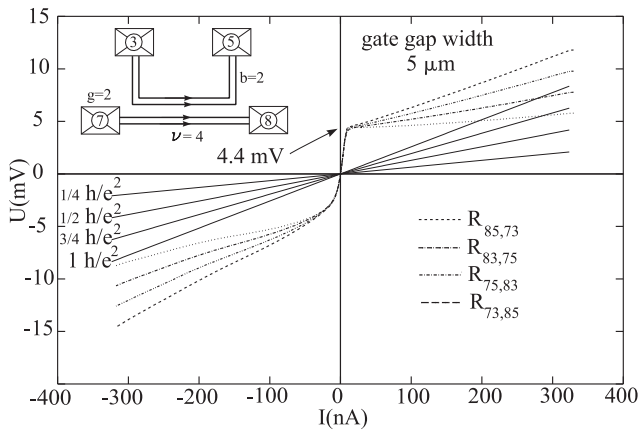} \vspace*{.3cm}
\caption{Equilibration across the Landau-gap, measured at 30 mK
and B = 3.89 T. The gate-gap width of the investigated sample is 5
$\mu$m and the chosen filling factors are $\nu=4$ and $g=2$. Solid
lines indicate the calculated 4-point resistances for the
different contact configurations according to the
Landauer-B\"uttiker-type formulas (equations ~\ref{eq:Req} and
~\ref{eq:Tfull}). The onset-voltage was determined to be 4.4 mV.}
\label{fig4}
\end{figure}
\end{center}

 Figure~\ref{fig4} shows I--V-traces
of a sample with 5 $\mu$m gate-gap width, obtained for the four
different contact configurations and filling factors $\nu=4$ in
the ungated areas and $g=2$ under the gate-electrode. As indicated
in the inset, this filling factor combination corresponds to
equilibration across the Landau-gap.

\vspace*{-.3cm}

The current-voltage characteristics are strongly nonlinear and
asymmetric. Regarding the low positive bias region in
Fig.~\ref{fig4}, which corresponds to a decrease of the
electrochemical potential of the two outermost edge channels in
the gate-gap, the occurrence of an onset-voltage is observable for
all contact configurations. At biases exceeding this voltage, the
experimental curves have an almost constant differential
resistance. For negative applied biases, no precise identification
of an onset-voltage is possible. Here, in contrast to what is
observed for positive biases, the I--V-traces are not exactly
reproducible from cooling to cooling and differ qualitatively from
sample to sample.

 At a positive current of about 10 nA, the slope of the traces shown
 in Fig.~\ref{fig4} decreases drastically and the differential resistances drop to
values close to those obtained for complete equilibration. We
interpret this step in the I--V-characteristic as a novel type of
breakdown-mechanism in the transport between adjacent edge
channels, which is not related to the usual, complete breakdown of
the quantum Hall effect. The latter is governed by transport
through the bulk of the 2DEG, which for macroscopic samples (like
the ones investigated here with lateral dimensions of a few
millimeters), is only observed at much higher currents
\cite{Shashkin94}. In particular, no step at small voltages is
found for a reference-sample with gate-gap widths zero, where only
the breakdown through the bulk can be seen, however at much higher
voltages. As discussed above, already at 4.2 K and even integer
filling factors the 2D bulk is in its insulating state and no
indications of breakdown are observed in the range of biases and
voltages considered here (see Fig.~\ref{fig3}). Also, bad Ohmic
contacts cannot account for the step in the I--V-characteristic as
the number of samples and contact combinations investigated make
this explanation highly unlikely. Furthermore, in the applied
4-terminal configuration, contact resistances should not
appreciably affect the results as long as the input-impedance of
the experimental setup is sufficiently high.

\subsection{Qualitative interpretation in terms of the
incompressible/compressible liquid model}\label{sec:Shk}

In this section we will develop a first interpretation of the
experimental observations shown in Fig.~\ref{fig4}, which is based
on the breakdown across the incompressible strip
\cite{Chklovskii93} between adjacent, but differently biased edge
channels. According to this picture it can also be understood, why
the onset-voltage of about 4.4 mV, roughly corresponds to the
Landau-gap $\hbar\omega_{c}$ for the given magnetic field $B=3.89$
T. Furthermore, the proposed model accounts for the asymmetry
between positive and negative current directions seen e.g. in
Fig.~\ref{fig4} and tries to explain why, at sufficiently large
currents the linear parts of the observed traces have almost the
same slope as the fully equilibrated resistance curves (see the
solid lines in Fig.~\ref{fig3} and the dash-dotted lines in
Fig.~\ref{fig6}, obtained from equations \ref{eq:Req} and
\ref{eq:leq}).

\vspace*{-.3cm}
\begin{center}
\begin{figure}
\epsfsize=0.74 \columnwidth \epsfbox{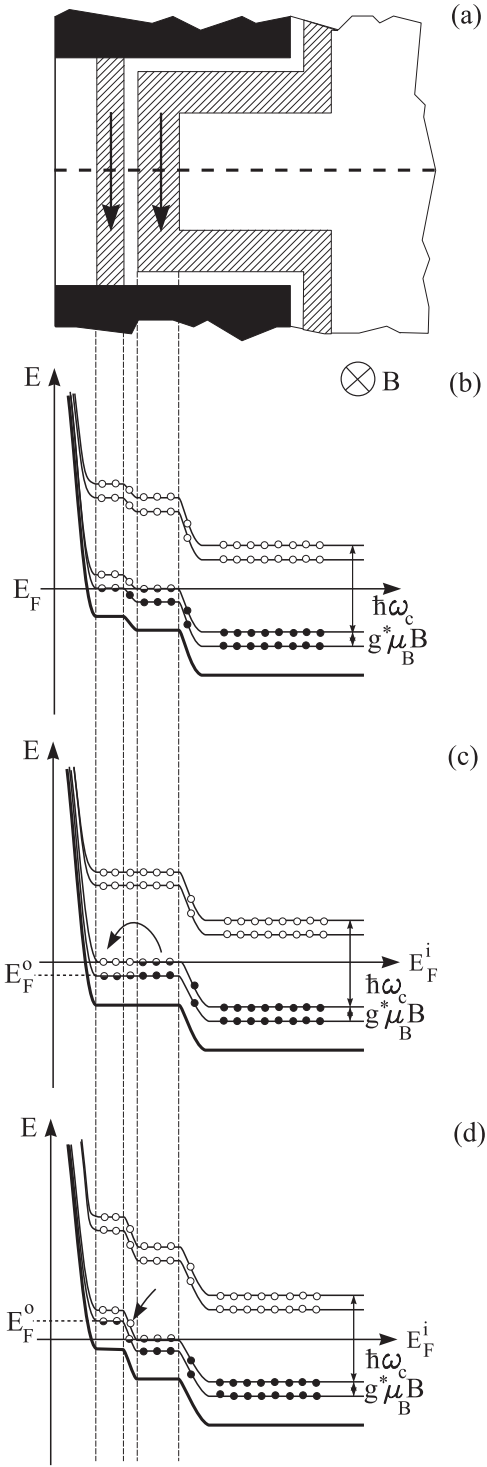} \vspace*{.1cm}
\caption{Application of the edge-reconstruction model to the
experimental setup. (a) Gate-gap region with compressible (shaded)
and incompressible (light) liquid strips. (b) Edge potential as
derived from the compressible/incompressible-liquid model in case
of a constant Fermi-energy in the gate-gap region. Filled circles
indicate occupied states, light ones unoccupied states. (c)
Shifted potential for an applied voltage $\Delta U_{oi}= + g^{*}
\mu_{B} B$ between two reservoirs at the inner and outer mesa
edge. (d) Edge potential for an applied bias of $\Delta U_{oi}= -
g^{*} \mu_{B} B$.} \label{fig5}
\end{figure}
\end{center}

The measured I--V-spectra in Fig.~\ref{fig4} resemble those of a
backward diode \cite{Sze81}. Indeed, in the picture of
edge-state-reconstruction developed by Chklovskii, Shklovskii et
al. \cite{Chklovskii92} (taking into account screening-effects
among electrons), transport across an incompressible edge-strip
can be described in a manner comparable to the transport across
the depletion-layer in a $p-n$ backward-diode. This is illustrated
in Fig.~\ref{fig5} (b)--(d), where we have sketched the edge state
reconstruction for $\nu= 2$, $g=1$. Also shown is the course of
edge channels in the gate-gap region for this filling factor
combination (Fig.~\ref{fig5} (a)). The mesa edge is positioned on
the left hand side, shaded areas indicate compressible liquids,
light regions incompressible strips and the bulk. The gate is
drawn in black. Arrows indicate the direction of electron drift in
the compressible liquid strips.

Figures \ref{fig5} (b)--(d) show the reconstruction of the edge
potential for the section indicated by the dashed line in (a).
Occupied states beneath the Fermi-energy are symbolized by full
circles, open circles represent unoccupied states, states at the
Fermi-energy are shown as half-filled circles. In the
configuration outlined in (b) no bias is applied between inner and
outer contacts and no current will flow between the separately
contacted edge states.

We now consider the case that one of the inner contacts is
grounded and a positive voltage is applied to an outer contact. In
this situation, the outer edge channel is shifted downwards in
energy with respect to the inner one. For a very low bias, this
aligns the occupied states in the inner channel with the
energy-gap in the outer channel so that ideally no current is
expected to flow. This gives reason for the appearance of the high
resistance region at very small positive bias in complete analogy
to the case of a backward-diode. Only when a high enough bias is
applied so that the topmost occupied inner edge state becomes
aligned with the first unoccupied outer edge state, transfer of
electrons between the edge channels becomes energetically allowed
and equilibration is readily achieved as depicted in
Fig.~\ref{fig5} (c). This explains the sudden breakdown of the
high differential resistance state when a positive voltage $\Delta
U_{oi}=+E_{g}/e$ is applied between inner and outer contacts.
Here, $E_{g}$ is the relevant energy-gap, which, for the case
$\nu=2$, $g=1$ is expected to be the spin-gap $g^{*}\mu_{B}B$,
with $g^{*}$ being the effective Land\'e-factor.

\vspace*{-0.3cm}
\begin{center}
\begin{figure}
\epsfsize=0.97 \columnwidth \epsfbox{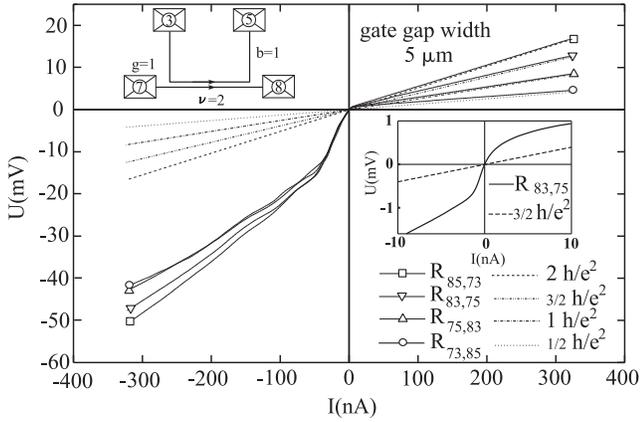} \vspace*{.2cm}
\caption{Filling factor combination $\nu=2$, $g=1$ measured at 30
mK and B= 7.75 T. Equilibration between inner and outer edge
channels occurs across the spin-gap in the first Landau-level.
Dotted lines indicate the calculated resistances for the case of
full equilibration. The inset shows a magnification of the
low-bias region taken for one contact combination at a low sweep
rate.} \label{fig6}
\end{figure}
\end{center}

For a negative bias, on the other hand, the first unoccupied inner
edge strip is energetically separated from the outer occupied
states by roughly the Landau-gap $\hbar \omega_{c}$ (see
Fig.~\ref{fig5} (d)). Therefore, a pronounced asymmetry is
expected in the I--V-spectra, as it is observed  experimentally
for $\nu=2$, $g=1$ (Fig.~\ref{fig6}).

 Here, at positive voltages
exceeding 1 mV, almost full equilibration is established, whereas
in the negative bias regime the onset of equilibration with a
higher transmission coefficient can be detected only for $|U|
> 10$ mV. It should also be pointed out, that for a negative bias no true
energy-gap opens up, since there are always unoccupied levels
(arrow in Fig.~\ref{fig5} (d)) at the quasi-Fermi-level of the
outer edge channels $E_{F}^{o}$. Therefore, no clear step-like
features are observed for negative biases.

 Of course this model
can directly be transferred to the case of equilibration across
the Landau-gap ($E_{g} = \hbar \omega_{c}$), as e.g. depicted in
Fig.~\ref{fig4}, where the step-like features are far more
pronounced and the asymmetry is still observable.

\subsection{Evaluation of the I--V-spectra}\label{sec:evaluation}

The characteristic onset-voltage for equilibration across the
Landau-gap is determined by linear extrapolation of the two
different curve branches in the positive voltage region. For the
data shown in Fig.~\ref{fig4}, we obtain $E_{g} = 4.4$ meV. A
comparison with $\hbar \omega_{c}$ for $B = 3.89$ T and an
effective electron-mass of $m^{*} = 0.067  m_{e}$ shows a
discrepancy of 2--3 meV. Similar deviations were observed for any
of the four investigated samples and both filling factor
combinations $\nu = 4$, $g = 3$ and $\nu = 4$, $g = 2$.

\vspace*{-0.3cm}
\begin{center}
\begin{figure}
\epsfsize=0.97 \columnwidth \epsfbox{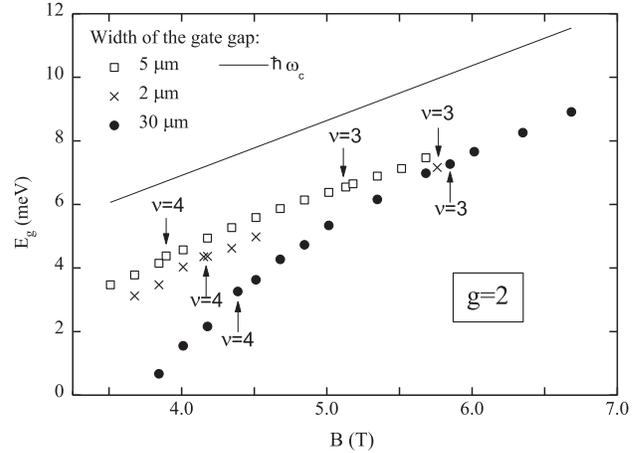} \vspace*{.1cm}
\caption{Magnetic field dependence of the determined energy-gaps
measured at 30 mK for different gate-gap widths. The filling
factor $g=2$ under the gate was held constant. The solid line is
given by the theoretical value of the Landau-gap $\hbar
\omega_{c}$.} \label{fig7}
\end{figure}
\end{center}

Figure~\ref{fig7} shows the magnetic field dependence of the
energy-gap determined for three different gate-gap widths. While
the filling factor $g=2$ under the gate is held constant, the
filling factor $\nu$ in the ungated areas varies with the strength
of the magnetic field. Note that even though at non-integer
filling factors the ungated bulk region are in a conducting state,
\cite{Chklovskii93} I--V-spectroscopy across the incompressible
strip is still possible. The solid line indicates the calculated
values of $\hbar \omega_{c}$. For $\nu \leq 4$ we obtain a linear
dependence for the samples with 2 and 5 $\mu$m gate-gap widths,
but again the above mentioned deviation from the theoretical value
of the Landau-gap is observed. A linearization of the curve slopes
--- where possible --- allows to determine the effective mass
$m^{*}$. The obtained values, as given in Table \ref{tab1}, are in
good accordance with the standard value of 0.067 $m_{e}$ for bulk
GaAs.

\begin{table}
\caption{Effective masses $m^{*}$ in units of the free electron
mass as calculated from the slopes of the linear fits to the data
shown in Fig.~\ref{fig7}. \label{tab1}} \vspace{0.3cm}
\begin{tabular}{lcr}
gate-gap width&$m^{*}/m_{e}$\\ \tableline 2 $\mu$m& 0.067 \\5
$\mu$m& 0.069 \\30 $\mu$m&0.060\end{tabular}
\end{table}

 The deviation from $\hbar \omega_{c}$ of roughly 2 meV can
partly be understood in terms of the edge-state model discussed in
section \ref{sec:Shk}, as the effective energy-gap width is
reduced by the spin-splitting energy $g^{*} \mu_{B} B$. The
Landau-level broadening might be a further cause for the onset of
equilibration at biases less than $\hbar \omega_{c}/e$. The
breakdown of the adiabatic transport regime at energy values
smaller but comparable to the Landau-level spacing was also
reported by Komiyama et al.\cite{Komiyama92,Machida96} for samples
with a cross-gate geometry. They proposed a self-consistent
reconstruction of edge states to describe the observed nonlinear
behaviour, finding $\hbar \omega_{c}/2$ to be the critical
potential-difference for nonequilibrium population of edge
channels.

 The evaluation of the data shown in Fig.~\ref{fig6} gives an effective
spin-gap of 0.57 meV. This corresponds to $g^{*} \approx 1.3$,
about three times the bulk GaAs $g$-factor. Because of the
smallness of the spin-gap it can only be evaluated from the
present data with a large margin of error. A survey of values
obtained for different gate-gap widths and filling factors (cf.
Fig. \ref{fig2}, top row) shows a large scatter, with values from
$g^{*} = 0.27$ to $g^{*} = 1.8$ \cite{Wuertz01}. A more thorough
investigation of the parameter-space with further high-resolution
measurements is therefore necessary.

 Assuming a constant equililibration-rate per unit-length, the
  equilibration-length $l_{eq}$ can be deduced from the linear part
of the I--V-curves at low positive voltages. This was done e.g.
for the filling factor combination $\nu = 4$, $g = 2$. Making use
of equation \ref{eq:Req} and \ref{eq:leq} we obtain $l_{eq}=$250
$\mu$m for 5 $\mu$m gate-gap width and 30 mK temperature. This
corresponds to a transmission-coefficient of T = 0.072. Similar
macroscopic equilibration-lengths have been reported by several
authors using different approaches
\cite{Muller92,Alphenaar90,Komiyama92,Hirai95}. Depending on the
gate-gap width we found equilibration-lengths up to the order of
magnitude of the sample size. Surprisingly the
equilibration-length also reveals a dependence on the choice of
contacts, which is not compatible with the above discussed
Landauer-B\"uttiker-type formulas.

\section{conclusion}

In summary, we have realized a pseudo-Corbino geometry, suitable
to obtain information on the energetic structure of the edge
potential in the quantum Hall regime by means of simple
I--V-spectroscopy. We directly observed non-linear transport
across different energy-gaps and determined the
transmission-coefficient for weak coupling among separately
contacted edge channels. Interpreting our results in terms of the
Landauer-B\"uttiker formalism \cite{Buttiker85,Buttiker88} and the
edge-reconstruction picture developed by Chklovskii et al.
\cite{Chklovskii92,Chklovskii93}, we show that the described
geometry enables us to shift the electrochemical potentials of
single edge states with respect to each other. Transport across
the incompressible liquid-strip can thereby be studied as a
function of the bias applied between inner and outer contacts.
Identifying the onset-voltage for inter-edge-channel current flow
with the energy-gap, we found for equilibration across the
Landau-gap a clear and reproducible deviation from the value of
$\hbar \omega_{c}$. On the other hand, the dependence of the
energy-gap on the magnetic field agrees well with the effective
mass of GaAs. Furthermore, from the observed spin-gap, an
effective $g$--factor can be determined, so far, however, only
with a large margin of error.

As mentioned, our new spectroscopic technique has addressed a
number of interesting questions regarding transport across
incompressible liquid-strips and the energetic structure of the
edge potential. Subject of future investigations with the proposed
sample geometry will --- besides continuing the study of
non-linear transport --- surely be the edge structure in the
fractional quantum Hall regime as well as the influence of
spin-flip processes on equilibration.

\acknowledgments

We wish to thank J.P. Kotthaus for his constant support of this
work and gratefully acknowledge financial support by the Deutsche
Forschungsgemeinschaft, SPP ``Quantum Hall Systems", under grant
LO 705/1-1. We also gratefully acknowledge help during the
experiments and discussions with A.A.~Shashkin. The part of the
work performed in Russia was supported by RFBR grant 00-02-17294,
INTAS YSF002, the programs ``Nanostructures" and ``Statistical
Physics" from the Russian Ministry of Sciences.

\appendix
\section{}

Referring to the geometry shown in Fig.~\ref{fig1} we calculate as
an example the 4-point resistance $R_{85,73}$. Again the
previously applied nomenclature is used: \bigskip

\begin{tabular}{lcl}
$\nu$ &:& filling factor in the ungated regions of the sample\\
$g$&:& filling factor in the gated regions \bigskip
\end{tabular}

From the application of the
Landauer-B\"uttiker-multichannel-multiprobe formula
\cite{Buttiker88} to the present geometry we derive the following
``current-statistics" for the 8 shown contacts:

\begin{eqnarray}
I_{1}&=&-\frac{e^{2}}{h}\nu \left(U_{1}-U_{2}\right)\nonumber\\
I_{2}&=&-\frac{e^{2}}{h}\left[\nu U_{2}-(\nu - g) U_{1}-g
U_{8}\right]\nonumber\\
 I_{3}&=&-\frac{e^{2}}{h} \nu
\left(U_{3}-U_{4}\right)\nonumber \\ I_{4}&=&-\frac{e^{2}}{h} \nu
\left(U_{4}-U_{5}\right)\nonumber\\ I_{5}&=&-\frac{e^{2}}{h} \nu
\left(U_{5}-U_{6}\right)\nonumber\\
I_{6}&=&-\frac{e^{2}}{h}\left[\nu U_{6}-g U_{3}- T_{36}
U_{3}-T_{76} U_{7} \right]\nonumber\\ I_{7}&=&-\frac{e^{2}}{h} g
\left(U_{7}-U_{1}\right)\nonumber\\
I_{8}&=&-\frac{e^{2}}{h}\left[g U_{8}-T_{78} U_{7}-T_{38} U_{3}
\right)\nonumber
\end{eqnarray}

with $T_{ij}$ being the transmission-coefficient for
charge-transfer from contact $i$ to contact $j$. Taking into
account charge conservation in the gate-gap region and assuming at
each contact the same number of incoming and outgoing edge
channels we further obtain:

\begin{eqnarray}
T &=& T_{38}= T_{76}\nonumber\\ g &=& T_{78}+ T_{38}\nonumber\\
\nu &=& g + T_{36}+ T_{76}\nonumber
\end{eqnarray}

Regarding now the 4-point resistance $R_{85,73}$, i.e. current
transport between the contacts 8 and 5, with contacts 7 and 3
being used as voltage probes, we receive a surprisingly simple
relation. (The currents at unused --- respectively floating ---
contacts are vanishing.)

\begin{eqnarray}
I_{1}&=& 0 \Rightarrow U_{1}= U_{2}\nonumber\\ I_{2}&=& 0
\Rightarrow U_{1}= U_{8}\nonumber\\ I_{3}&=& 0 \Rightarrow U_{3}=
U_{4}\nonumber\\I_{4}&=& 0 \Rightarrow U_{4}=
U_{5}\nonumber\\I_{6}&=& 0 \nonumber\\ I_{7}&=& 0 \Rightarrow
U_{1}= U_{7}\nonumber
\end{eqnarray}

\begin{eqnarray}
R_{85,73}\cdot \frac{e^{2}}{h}&=&\frac{U_{3}-
U_{7}}{I_{8}}\nonumber\\&=&\frac{U_{7}-U_{3}}{g U_{8}- \left(g -
T_{38}\right) U_{7} - T_{38} U_{3}}\nonumber\\&=& \frac{U_{7} -
U_{3}}{T_{38}\left(U_{7} - U_{3} \right)} =
\frac{1}{T_{38}}\nonumber\\&=& \frac{1}{T}\nonumber
\end{eqnarray}


\begin{references}
\bibitem{byline} The data shown in Fig.~\ref{fig3} were taken
on a sample with a slightly different geometry, having two
separate inner etched regions. In this case, the fully
equilibrated resistances in units of $h/e^2$ are 1/4, 1/2, 3/4,
and 1 for the configurations $R_{73,85}$, $R_{83,75}$, $R_{75,83}$
and $R_{85,73}$, respectively.
\end{references}
\end{document}